\documentclass[showpacs,preprint]{revtex4}
\usepackage{graphicx}% Include figure files
\usepackage{dcolumn}% Align table columns on decimal point
\usepackage{bm}% bold math

\begin{document}

\title{Non-Life Insurance Pricing : Statistical Mechanics Viewpoint}
\author{Amir H. Darooneh}
\affiliation{Department of Physics, Zanjan University, P.O.Box
45196-313, Zanjan, Iran.} \email{darooneh@mail.znu.ac.ir}
\date{\today}

\begin{abstract}
We consider the insurance company as a physical system which is
immersed in its environment (the financial market). The insurer
company interacts with the market by exchanging the money through
the payments for loss claims and receiving the premium. Here in
the equilibrium state we obtain the premium by using the canonical
ensemble theory, and compare it with the {\it Esscher} principle,
the  well known formula in actuary for premium calculation. We
simulate the case of car insurance for quantitative comparison.
\end{abstract}

\pacs{89.65.Gh, 05.20.-y}
 \maketitle

\section{Introduction}

From the first of nineteen century the economists have tried to
find a way to use the formalism of physical theories in
mathematical modeling of economy. The concepts of utility function
and economical equilibrium entered into the financial theories in
correspondence with potential energy and mechanical stability
\cite{bjn}. The physicists have paid less attention to this
subject till last decade \cite{bjn}. Recently they have more
regards for studying the dynamics of the stock
markets\cite{bp,ms}. The main source of their interest comes from
fluctuation of stock prices with the time. Understanding this
behavior enable us to manage the strategies for trading \cite{sz}.
The statistical mechanics has appeared as a powerful method for
exploring the price dynamics \cite{bjn,bp,ms,sz}.

The insurance is one of the important parts of the financial
market with respect to trading risks. In the specified period of
time, the results of a loss events (risks) that occur for the
insurant are covered by insurance company and insurant also pays
an amount of money to the insurer, which is called premium
\cite{gv,kpw}. In the competitive financial market the calculation
of premium is very complicated. The premium is affected by random
nature of risks and also variation in number of the insurants. The
latter may be decreased due to an increment in premium and is
increased by reduction of premium. The actuary is a branch of
mathematics that studies the relation between paid and received
money in order to assign premium to a category of risks.

Nowadays the physicists also pay attention to this part of the
financial market. They look for a new way for the premium
calculation on the basis of physical concepts \cite{fd,d}.

In this paper we are intended to explain how the equilibrium
statistical mechanics may be used for premium forecasting. In the
next section we describe the analogy between the financial market
and physical system in contact to heat reservoir. A method for
equilibrium insurance pricing will be suggested based on the
canonical ensemble theory. In the last section the special case of
car insurance is simulated and we compare our method with {\it
Esscher} principle which is the famous method in actuary \cite{gv}
and asset pricing \cite{gs,b2,b3,vgdkd}.

\section{Description of the Model}

The financial market is combination of large number of economic
agents which are interacting with each other through buying and
selling. We consider the behavior of one of the agents for example
an insurance company; all other agents may be regarded as its
environment. The agent exchanges money when interacts with its
environment. We suppose the financial market is a closed system,
the environment will absorb the money that the agent loses and
supplies the agent's incomes. In equilibrium state a parameter
that we denote it by $\beta$ has the same value for the agent and
its environment \cite{p}.

\begin{equation}
\beta_{Agent}=\beta_{Environment}
\end{equation}

This parameter should be related to macroscopic properties of an
economic agent. Later on, we signify it for an insurance company
in terms of its initial wealth, the mean claim size and the ruin
probability.

Like a physical system in contact with a heat reservoir we can
nominate the following probability for $E(t)$, the net money which
is exchanged between an agent and its environment at the end of
specified time interval \cite{p},
\begin{equation}
  Pr(E(t))=\frac{e^{-\beta E(t)}}{\sum e^{-\beta E(t)}}
\label{e1}
\end{equation}
The sum is over all possible amounts of exchanged money in
equilibrium state. We assume here that the agent's money is
considerably lesser than the money in the environment.

The above result is confirmed by simulation \cite{cc,dy1} and is
also inferred from empirical data \cite{dy2,dy3}.

For the special case of car insurance in the time interval $t$,
the insurance company receives premium from $I(t)$ insurants and
covers the loss results for $N(t)$ car accidents. In each accident
insurant charges the company for $X_{j}$ amount of money. Thus the
exchanged money at the end of interval is,

\begin{equation}
  E(t)=pI(t)-\sum^{N(t)}_{j=1}X_{j}
\label{e2}
\end{equation}
Where $p$ is the premium. In the insurance terminology we call
this money the surplus of company. The $I(t)$, $N(t)$ and $X_{j}$
are random variables which their probability distribution should
be inferred from empirical data of the insurance company.

Premium calculation principle is a rule that assign to any
distribution function (correspond with loss events) a real number
\cite{gv}. B\"{u}hlmann in his brilliant papers on economic
premium principle stated that premium calculation principle should
also depend on market conditions in addition to loss distribution
\cite{b4,b5}. He constructed a model for whole insurance market
with many agents; like insurance companies, reinsurance companies
and buyers of insurance. He concluded that the sum of all incurred
losses for all agents in the market shows the effect of
surrounding market and it should be included in premium
calculation principle.

In this work we introduce a new way for premium calculation. The
number of insurants enters in our model as a reminiscence of the
market conditions, the competition in the market changes it
randomly as we mentioned above.

The insurance company avoids of financial failure. Premium should
be calculated so that the surplus of company at the end of
specified interval should be zero at least. This condition may be
satisfied in mean.
\begin{equation}
<E(t)>=\sum Pr(E(t))E(t)=0.
 \label{e3}
\end{equation}

In real case premium is greater than the value obtained from
eq.~\ref{e3} because administration expenditures must be
compensated and company also likes to profit, anyhow we aren't
considering these matters.

If the number of insurants becomes constant then by simple
calculation eq.~\ref{e3} reduces to what is known as the {\it
Esscher} principle.

\begin{equation}
pI_0=\frac{\sum Z(t)e^{\beta Z(t)}}{\sum e^{\beta Z(t)}}.
\label{e4}
\end{equation}
Where $Z(t)$ is the aggregate loss in the time interval $t$, it is
defined as,
\begin{equation}
 Z(t)=\sum_{j=1}^{N(t)}X_j
\end{equation}

Unlike to the net premium that is obtained by averaging the
aggregate loss, in the \textit{Esscher} principle the premium is
calculated by exponential tilting of loss events. The parameter
$\beta$ appears to scale the weight of each event. There is no
unique way for computing the $\beta$ parameter in \textit{Esscher}
principle; some authors use the optimization procedure \cite{gv}
and others relate it to the risk tolerance of the insurance
company \cite{b4}. The way that we adopt in this paper for
determination of this parameter is also applicable here.

The \textit{Esscher} principle as an efficient method for premium
computing appears as a special case of our method. The difference
between it and our method is shown in the next section
numerically.

%In actuary literature the parameter $\beta$ in eq.~\ref{e4} is
%obtained by optimization procedure \cite{gv} or it can be related
%to the risk tolerance of the insurance company \cite{b4}. The way
%for determination of this parameter in our method which come in
%the next section may be also applied here.

\section{Simulation Results}

To simulate our model, it is necessary to specify the
distributions for the random quantities which are appeared in
expression for the surplus, eq.~\ref{e1}.

By assumption the number of loss claims by insurants, $N(t)$ has
Poisson distribution with the mean $\lambda t$. It depends only on
the length of interval $t$ and also is statistically independent
of the number of claims in any previous interval.
\begin{equation}
 Pr(N(t+s)-N(s)=n)=\frac{{(\lambda t)}^n e^{-\lambda t}}
 {n!}
\end{equation}

The important property of Poisson distribution is that the times
between successive claims are independent and identically
exponentially distributed (the {\it Erlang}'s model) with the mean
$1/\lambda$. It is what we encounter in most of the real cases.

The model for the claim size is constructed based on the knowledge
and experience of the insurance company in addition to data from
the past. The model should provide a balance between simplicity
and conformity to the available data.

The purpose of this paper is explanation of our method; in this
respect for simplicity we restrict ourselves to the exponential
distribution for the claim size. Otherwise we are forced to do a
lengthy numerical computation that don't correspond to our goal.

\begin{equation}
Pr(X=\xi)=\frac{1}{\mu}e^{-\xi/\mu}
\end{equation}

The $\mu$ is the mean of the claim size. The product of the
$\lambda t$ and $\mu$ is the net premium, $p_0$. The insurance
company which offer this value or less for the premium will be
busted ultimately \cite{kpw}.

The distribution for the number of insurants has not been modeled
since now because it doesn't be involved in the traditional
methods for calculating the premium. The eq.~\ref{e3} shows its
importance in our method. However we assume the number of
insurants is distributed uniformly with the mean, $I_0$, and
standard deviation, $\%10 I_0$. This assumption based on empirical
data of the Iran Insurance Company \cite{r}.

All the necessary parameters for tuning the mentioned distribution
functions are obtained from statistical reports of Iran Insurance
Company for the case of car insurance in year 2000 \cite{r}. For
independence of the results to any monetary units, the initial
wealth and premium are expressed in terms of the mean claim size
and net premium. The following results are averaged over 10000
different possible states of the modeled insurance company.

\begin{figure}
\includegraphics[width=8.5cm,height=5.5cm]{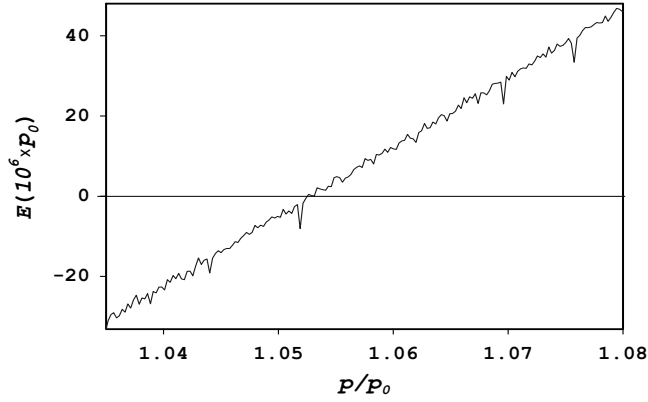}
\caption{\label{fig1} The annual profit or surplus after $300$
days versus the premium for large $\beta$. The premium and surplus
are measured in terms of net premium.}
\end{figure}

The variation of annual profit or surplus after $300$ days with
premium for large $\beta$, is shown in the fig.~\ref{fig1}. The
desired value for the premium is obtained from calculation of the
intersection point of approximated curve with the premium axis.
For other values of the parameter $\beta$ we can apply the same
method for computing the premium.

\begin{figure}
\includegraphics[width=8.5cm,height=5.5cm]{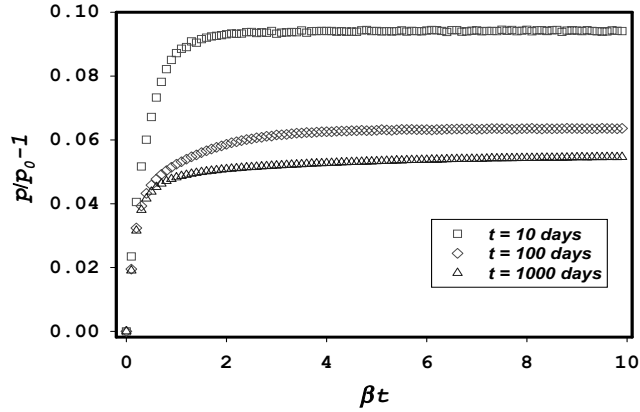}
\caption{\label{fig2}The premium as a function of $\beta t$. The
results for $t=10,100$ and $1000$ is plotted.}
\end{figure}

The premium is also function of parameter $\beta $. In the case of
$\beta=0$ it is equal to the net premium. An increment in $\beta$
increases the premium but for large $\beta$ it approaches to a
limiting value. The fig.~\ref{fig2} shows this behavior for three
value of the time interval.

\begin{figure}
\includegraphics[width=8.5cm,height=5.5cm]{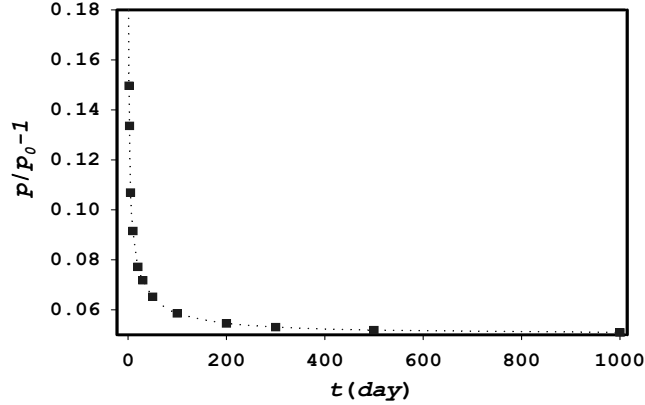}
\caption{\label{fig3}The premium versus duration of the insurance
contract for the large $\beta$. The squares are obtained by
simulation and the curve shows the fitting result.}
\end{figure}

The fig.~\ref{fig3} demonstrates dependency of the premium on
duration of insurance contract for the large $\beta$. This
behavior is expected in the insurance market \cite{kpw}. The
premium for other values of the parameter $\beta$ shows the same
behavior in the time.

\begin{figure}
\includegraphics[width=8.5cm,height=5.5cm]{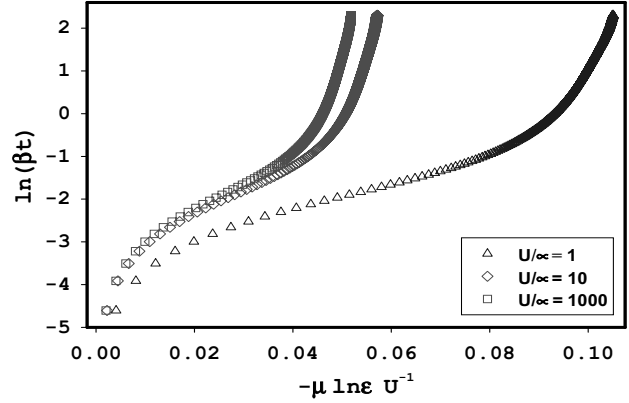}
\caption{\label{fig4}The parameter $\ln(\beta t)$ as a function of
$\mu ln\varepsilon U^{-1}$ which is introduced in the text. The
results for $U=1,10$ and $1000$ times of the mean claim size are
plotted. }
\end{figure}

The parameter $\beta$ has been ambiguous since now. At this moment
we proceed to clear its meaning. It can be understood by
dimensional analysis of the eqs.~\ref{e2} and ~\ref{e3} that the
parameter $\beta$ is proportional to the inverse of time. The
fig.~\ref{fig2} also confirms this statement, the behavior of the
premium as a function of $\beta t$ is the same for different time
intervals in exception of their limiting values.

The most important quantities which are used by the insurers for
their financial decisions are the initial wealth $U$ and the
ultimate ruin probability $\varepsilon$. The combination of them
with the mean claim size $\mu$ is used in all methods for
calculating the premium \cite{gv,kpw}. The parameter $\beta$ is
also function of these quantities. The fig~\ref{fig4} shows the
$\ln\beta t$ as a function of the dimensionless parameter
$-\mu\ln\varepsilon U^{-1}$ and the initial wealth. This
consequence is issued from relation of the ultimate ruin
probability on premium \cite{kpw},
\begin{equation}
\varepsilon={p_0\over p}\exp{[-{U\over\mu}({p_0\over p}-1)]}.
\label{e5}
\end{equation}
And the result for dependency of premium on $\beta t$ which is
also shown in fig.~\ref{fig2}.

The ultimate ruin probability for exponential distribution of
claim size can be calculated analytically as is seen in the
eq.~\ref{e5}, but in other cases it would be computed numerically
from the following relation \cite{kpw},
\begin{equation}
 \varepsilon=1-\sum_{k=0}^{\infty}(1-{p_0\over p})({p_0\over
 p})^k F_e^{*k}(U).
\end{equation}
Where $F_e(x)$ is related to the claim size probability $F(x)$,
\begin{equation}
F_e(x)={1\over\mu}\int_0^x[1-F(y)]dy
 \end{equation}
And $F_e^{*k}(U)$ is the $k$-fold convolution of $F_e(U)$ with
itself.

The fig.~\ref{fig4} also indicates if the initial wealth becomes
large for a fixed time interval the parameter $\beta$ approaches
to zero. This means that a wealthy company buys the risks with the
less premium in comparison to the small companies which offer more
expensive insurance. This fact conforms to our experience in the
financial market.

\begin{figure}
\includegraphics[width=8.5cm,height=5.5cm]{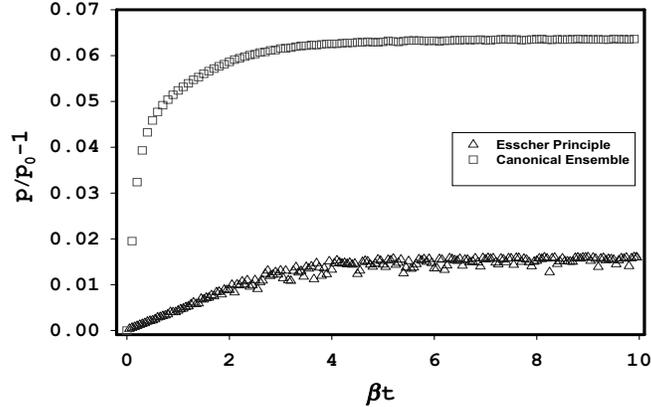}
\caption{\label{fig5}The premium versus the parameter $\beta t$
for $100$ days interval. The results of our method (canonical
ensemble theory) and {\it Esscher} principle are plotted for
comparison.}
\end{figure}

As mentioned before the {\it Esscher} principle is concluded from
our method. The variation in number of insurants changes the
premium certainly. The fig.~\ref{fig5} displays the difference
between two methods numerically.

\section{summary}
We construct a simple model for the cash flow in an insurance
company for the category of non-life insurance. A new method is
suggested for calculation of premium on the basis of the canonical
ensemble theory. This method considers the effect of the
environment (or market conditions) by introducing a new random
variable, the number of insurants, into the equation of cash flow.
In this respect the empirical data of an insurance company is
sufficient for take into consideration of environment conditions.
There is no need to acquire any information about other agents in
the market unlike the B\"{u}hlmann method. Another advantage of
this method is ability for extension in the small market case. The
latter is under investigation by the author.

The financial backbone of insurer company is concealed in the
parameter $\beta$ which play essential role in our method for
computing the premium. When we omit the effect of market
conditions by considering constant value for number of insurants
the {\it Esscher} principle is emerged as a result. Difference
between our method and the {\it Esscher} principle that is
apparently due to the effects of environment is shown numerically.

This work has been supported by the Zanjan university research
program on Non-Life Insurance Pricing No 8243.

\bibliographystyle{}

\begin{thebibliography}{}

\bibitem{bjn}Z. Burda, J. Jurkiewicz and M. A. Nowak, {\it Is Econophysics a Solid
Science?}, Acta Physica Polonica B, {\bf 34}, 87 (2003).

\bibitem{bp}J.P. Bouchaud and M. Potters, {\it Theory of Financial Risks},
Cambridge University Press, Cambridge, (2000).

\bibitem{ms}R. Mantegna and H.E. Stanely, {\it An Introduction to
Econophysics}, Cambridge University Press, Cambridge, (2000).

\bibitem{sz}D. Sornette and W. X. Zhou, {\it The US 2000-2002 market descent},
 Quant. Finance, {\bf 2}, 468 (2002).


%\bibitem{s}A. H. Sharif, {\it Premium Calculation by Transformed Distributions},
% Actuarial Research Clearing House, {\bf 1}, 111 (1996).

\bibitem{gv}M. J. Goovaerts and F. deVylder, {\it Insurance Premium},
 North Holland, Amsterdom, (1984).

\bibitem{kpw}S. A. Klugman, H. H. Panjer and G. E. Willmot, {\it Loss Models: From Data to Decisions},
 John Wiley, New York, (1998).

\bibitem{fd}M. E. Fouladvand and A. H. Darooneh,
{\it Premium Forecasting for an Insurance Company}, Second Nikkei
Workshop and Symposium on Econophysics (Ed. H. Takayasu), 313 ,
Springer, Berlin, (2003).

 \bibitem{d}A. H. Darooneh, {\it Non Life Insurance Pricing},
Proceeding of 18-th Annual Iranian Phys. Conf., 162 (2003). {\it
in persian}

\bibitem{gs}H. U. Gerber and Shiu E. S. W., {\it Option Pricing by Esscher Transforms},
 Transaction of the Society of Actuaries,{\bf XLVI}, 99 (1994).

\bibitem{b2}H. Buhlmann, F. Delbaen, P. Embrechts and A. Shirayev,
{\it No-Arbitrage Change of Measure and Conditional Esscher
Transform}, CWI Quarterly,{\bf 9}, 291 (1997).


 \bibitem{b3}H. Buhlmann, F. Delbaen, P. Embrechts and A. Shirayev, {\it On Esscher Transform in Descrete finance Model},
 ASTIN Bulletin,{\bf 28}, 171 (1998).

\bibitem{vgdkd}D. Vyncke, M. J. Goovaerts, A. DeSchepper, R. Kass and J. Dhaene,
{\it On the Distribution of Cash-Flows using Esscher Transforms},
 Journal of Risk Insurance,{\bf 70}, 563 (2003).

\bibitem{p}R. K. Pathria, {\it Statistical Mechanics},
 Pergamon Press, Oxford, (1972).

\bibitem{cc}A. Chatterjee and B. K. Chakarbarti, {\it Money in Gas-Like Market},
 Physica Scripta, {\bf T90}, 36 (2003).

\bibitem{dy1}A. Dr$\breve{a}$gulescu and V. M. Yakovenko, {\it Statistical Mechanics of Money},
 Eur. Phys. J. B, {\bf 17}, 723 (2000).

\bibitem{dy2}A. Dr$\breve{a}$gulescu and V. M. Yakovenko,
{\it Evidence for the Exponential Distribution of Incomes in the
USA },  Eur. Phys. J. B, {\bf 20}, 585 (2001).

\bibitem{dy3}A. Dr$\breve{a}$gulescu and V. M. Yakovenko,
{\it EExponential and Powe Law Probability Distribution of Wealth
and Incomes in the United Kingdom and United States }, Physica A,
{\bf 299}, 213 (2001).

 \bibitem{b4}H. Buhlmann, {\it An Economic  premium Principle},
 ASTIN Bulletin,{\bf 11}, 52 (1980).

 \bibitem{b5}H. Buhlmann, {\it General Economic  premium Principle},
 ASTIN Bulletin,{\bf 14}, 13 (1984).

 \bibitem{r}\text{-----} , {\it Statistical Report of Iranian Insurance Industry},
  Iran Central Insurance Company, Tehran, (2000).


\end{thebibliography}

\end{document}